\documentclass[a4paper,fleqn,usenatbib]{mnras}

\usepackage{newtxtext,newtxmath}

\usepackage[T1]{fontenc}
\usepackage{ae,aecompl}
\usepackage{siunitx}
\usepackage{xcolor}

\usepackage{graphicx}	
\usepackage{amsmath}	
\usepackage{amssymb}	
\usepackage{multirow}
\usepackage{booktabs}

\title[Wavefront prediction using ANNs for open-loop AO]{Wavefront prediction using artificial neural networks for open-loop Adaptive Optics}

\author[Xuewen Liu et al.]{Xuewen Liu,$^{1}$\thanks{E-mail: xuewen.liu@durham.ac.uk}
Tim Morris,$^{1}$
Chris Saunter,$^{1}$ 
Francisco Javier de Cos Juez,$^{2}$ 
\newauthor{Carlos Gonz\'{a}lez-Guti\'{e}rrez,$^{2}$
Lisa Bardou$^{1}$}
\\
$^{1}$Centre for Advanced Instrumentation, Department of Physics, Durham University, South Road, Durham DH1 3LE, UK \\
$^{2}$University Institute of Space Sciences and Technologies of Asturias, University of Oviedo, 33004 Oviedo, Spain
}

\date{Accepted XXX. Received YYY; in original form ZZZ}

\pubyear{2019}

\begin{document}
\label{firstpage}
\pagerange{\pageref{firstpage}--\pageref{lastpage}}
\maketitle

\begin{abstract}
Latency in the control loop of adaptive optics (AO) systems can severely limit performance. Under the frozen flow hypothesis linear predictive control techniques can overcome this, however identification and tracking of relevant turbulent parameters (such as wind speeds) is required for such parametric techniques. This can complicate practical implementations and introduce stability issues when encountering variable conditions. Here we present a nonlinear wavefront predictor using a Long Short-Term Memory (LSTM) artificial neural network (ANN) that assumes no prior knowledge of the atmosphere and thus requires no user input. The ANN is designed to predict the open-loop wavefront slope measurements of a Shack-Hartmann wavefront sensor (SH-WFS) one frame in advance to compensate for a single-frame delay in a simulated $7\times7$ single-conjugate adaptive optics (SCAO) system operating at 150 Hz. We describe how the training regime of the LSTM ANN affects prediction performance and show how the performance of the predictor varies under various guide star magnitudes. We show that the prediction remains stable when both wind speed and direction are varying. We then extend our approach to a more realistic two-frame latency system. AO system performance when using the LSTM predictor is enhanced for all simulated conditions with prediction errors within 19.9 to 40.0~nm RMS of a latency-free system operating under the same conditions compared to a bandwidth error of $78.3\pm4.4$~nm RMS.
\end{abstract}




\section{Introduction}
In adaptive optics (AO) systems, time lag between wavefront detection and correction induces the bandwidth error. For Extreme AO (XAO) systems for high contrast imaging (HCI) of exoplanets, the bandwidth error results in broadening of the point spread function (PSF) along dominant wind directions, which severely degrades contrast, especially at small star separations \citep{Kasper12, Males2018}. For wide-field AO systems dominated by tomographic errors, to keep bandwidth error tolerable, the integration time of wavefront sensing and thus guidable star magnitude (either natural or laser) is limited, which then limits the sky coverage \citep{Correia14,Jackson15}. One way to overcome this problem is to predict the future wavefront from recent past wavefront measurements. Under the frozen flow hypothesis \citep{Wang08,Poyneer09}, the turbulence volume is modeled as a linear composition of static, independent layers, each translating across the telescope aperture with certain velocity as a result of dominant wind at that layer. Because of this spatial and temporal correlation, it is possible that the future wavefronts can be partially predicted using past measurements. This hypothesis is a reasonable simplification of the turbulence for wavefront prediction purposes.

Predictive control in AO is an active research area that incorporates wavefront prediction based on the frozen flow hypothesis into controller design. One of the most popular schemes is the Kalman filter based Linear Quadratic Gaussian (LQG) control \citep{Paschall1993,LeRoux2004}. Under this framework, the whole system (both turbulence and AO system) is represented by a small set of state variables. Linear models are used to describe temporal evolution of those variables as well as their links with system measurements. Priors from system telemetry and noise statistics are then combined to obtain the control law. Because of its flexibility in structure, LQG predictors allow for additional consideration of other system error sources such as static error and vibration. Numerical and laboratory implementations focusing on a single or a few Zernike modes show great improvement in terms of overall residual phase error or Strehl ratio \citep{LeRoux2004,Kulcsar12} and especially vibration filtering \citep{Petit2006,Petit2008}. \citet{Poyneer07} developed a computationally efficient Fourier based LQG predictive controller, which can be extended to non-integer loop delays \citep{Poyneer08}, facilitating graceful formulation of wind-blown turbulence evolution under Fourier basis. Laboratory tests demonstrate a reduction of around 67\% in bandwidth error using a full Fourier LQG controller \citep{Rudy15}. \citet{Correia14} incorporates open-loop wavefront prediction into a minimum mean square error (MMSE) tomographic reconstructor design for multi-object AO systems. This tomographic predictor allows for use of one-magnitude fainter guide stars (corresponding to an increase in the density of available stars by a factor of 1.8) in end-to-end simulations of RAVEN \citep{Andersen2012}, which is expected to be further improved if deployed within a LQG framework. LQG based predictive control has been deployed for AO systems on HCI instrument SPHERE \citep{Petit2014} for both turbulence correction and vibration filtering in tip-tilt modes. Stability and robustness of LQG controller in full-mode single-conjugate AO (SCAO) control has also been verified on sky \citep{Sivo2014}, showing overall performance improvement over a standard integrator controller in conditions where bandwidth error is not dominant. 

The recently proposed Empirical Orthogonal Functions framework \citep{Guyon17} for predictive control aims at fully exploiting linear spatio-temporal correlations within input telemetry and improving controller robustness by assuming no physical model of turbulence evolution. Numerical HCI simulations demonstrate significantly improved contrast and robustness against sensor noise. Although this feature can significantly simplify practical implementation, frequent re-learning and update is unavoidable, for such data-driven predictor and above-mentioned LQG approach, to adapt to varying turbulence conditions. 

In this paper, we exploit the potential of artificial neural networks (ANNs) as a nonlinear framework for wavefront prediction. Early numerical simulations adopting a feed-forward multi-layer perceptron (MLP) network demonstrate promise for using this nonlinear tool for slope prediction based on a time series of past noisy measurements by a Shack-Hartmann wavefront sensor (SH-WFS) \citep{Jorgenson92,Jorgenson1994}, with further improvement over a linear predictor when signal-to-noise ratio (SNR) of wavefront sensing gets lower \citep{Hart1996}. The last few decades have seen significant advances in both the theory and applications of ANNs \citep{LeCun2015}, among which the Long Short-Term Memory (LSTM) network is well-suited to time series modeling and prediction by design \citep{Hochreiter1997,Gers1999}. 

\section{Artificial neural networks}
ANNs are computational models inspired by biological neural networks. They are composed of a series of computing elements called neurons that are interconnected in a layered structure. Each neuron receives inputs, either as an input to the entire network or as outputs of connected neurons in the former layer, then transmits mathematically processed input information to connected neurons in the next layer. This forward transmission continues until the final output neurons are reached. Information flow in each type of ANNs is thus specified. A thorough tutorial on ANNs can be found in \citet{Goodfellow2016}. A detailed description of a MLP network and its successful application for tomographic wavefront reconstruction can be found in \citet{Osborn2012}.  

Long Short-Term Memory (LSTM) is an advanced architecture of recurrent neural networks (RNNs) that are specially designed for processing sequential data \citep{Graves2012}. Compared with MLPs with forward transmissions only, RNNs have dynamic feedback connections and shared parameters across all time steps. An internal state vector is transferred through time to maintain memories. LSTMs can especially cope well with long-term dependencies \citep{Goodfellow2016}, which otherwise renders training in normal RNNs much more difficult. It has been successfully applied in fields such as speech recognition \citep{Graves2013}, machine translation \citep{Sutskever2014} and image captioning \citep{Karpathy2014}. LSTMs have two desirable features for wavefront prediction:
\begin{itemize} 
    \item \textbf{No user input}. No prior knowledge of the atmosphere is assumed for the training process. No user input is required either during application.
    \item \textbf{No user tuning}. The fluid nature of the memory elements within allows the network to learn temporal behaviours of turbulence of varied time constants and to adapt to changes in these without user tuning. The nonlinearity of LSTMs enables the agility and robustness when dealing with non-frozen flow turbulence evolution (such as fluctuations in wind velocities), WFS noise or change of turbulence strength.
\end{itemize}

\section{Methodology} 
We exploit the potential of ANNs for wavefront prediction in numerical simulations based on a SCAO system. More specifically, the ANN predictor is trained in simulation to predict uncorrected wavefront slopes at the next time step based on a time series of past noisy slope measurements by a SH-WFS operating in open loop. The simulated SCAO system serves two purposes. The wavefront sensing subsystem is used to generate a series of time sequences of wavefront slopes as training data, with the last frame of slopes in each sequence being the training target. After training, the predictor is incorporated into the AO correction loop for evaluation. 

To quantify the efficacy of the ANN predictor, we compare AO corrections in terms of root-mean-square wavefront errors (RMS WFE) under three operating conditions, depending on which WFS measurement is applied to DM at time step $t$: 
\begin{itemize}
\item Zero-delay or delay-compensated loop, where the current measurement $\textbf{s}_{t}$ is used immediately.
\item One-frame delay loop, where the prior measurement $\textbf{s}_{t-1}$ is used.
\item ANN predictive loop, where the predicted current measurement $\Tilde{\textbf{s}}_{t}$ from $(\textbf{s}_{1}, \textbf{s}_{2},...\ ,\textbf{s}_{t-1})$ is used.
\end{itemize}

\subsection{SCAO simulation}
\label{subsec:simDescription}
\begin{figure*}
\centering
\includegraphics[width=12cm]{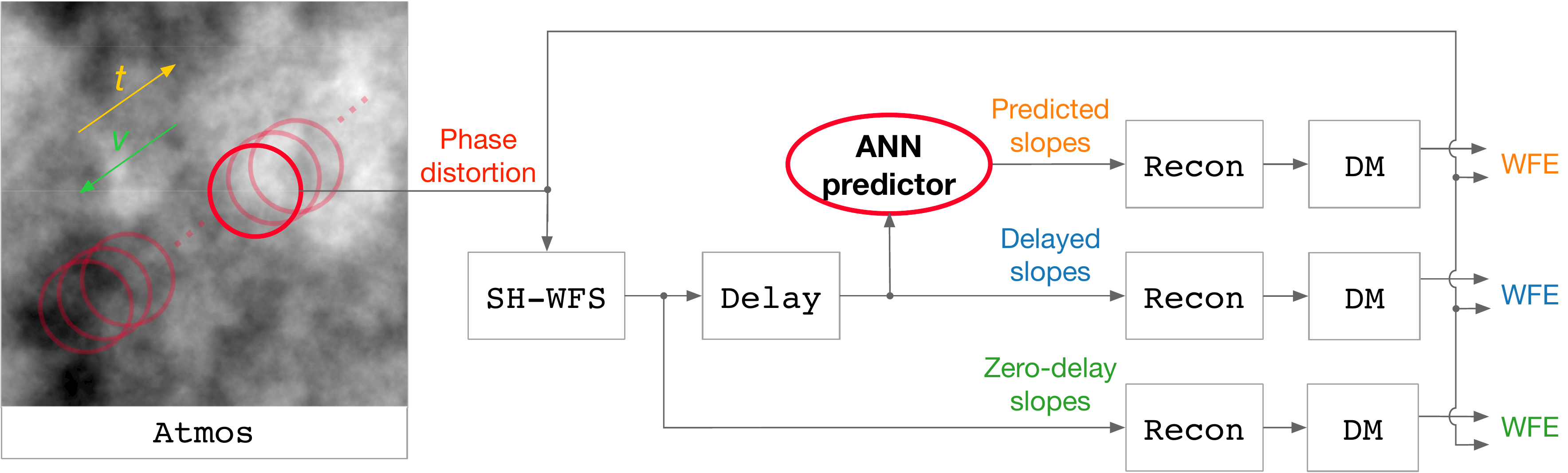}
\caption{Composition of the simulated SCAO system and its data flow. RMS wavefront error of the predictive correction (upper) is expected to be between the delayed (center) and delay-compensated (lower) corrections.} 
\label{fig:soapy}
\end{figure*}

The AO simulation tool used is Soapy (Simulation `Optique Adaptative' with Python) \citep{Reeves2016}. Soapy is highly modular, enabling both end-to-end simulations and fast experimentation using a subset of its modules. New modules can also be easily integrated. 
The architecture of the simulated SCAO system is shown in Fig.~\ref{fig:soapy}. Throughout the simulation we use a point source at infinity to act as a natural guide star (GS). To generate the training data, one single turbulence layer is assumed. Here, the use of a single layer is for the ease of training. We will show that a ANN predictor trained using one turbulence layer is capable of predicting in multi-layer conditions. A large random phase screen with Von Karman statistics is generated within the atmosphere module \texttt{Atmos} at the start of each loop run. Pure frozen flow is assumed, under which the large phase screen is translated over the telescope aperture with a given velocity due to the wind. At each time step, a smaller portion of the large phase screen, the part of which is seen by the telescope aperture, is output to \texttt{SH-WFS}. \texttt{SH-WFS} then outputs measured noisy wavefront slopes from the image plane using thresholding centre of gravity (TCoG). The thresholding value is a flux cutoff described by a factor of the maximum intensity within a subaperture to suppress photon noise and readout noise. A single frame delay can be used in Soapy simulations (the center loop in Fig.~\ref{fig:soapy}) to account for the inevitable WFS integration time. This time lag between wavefront measurement and correction can be compensated either by applying slope measurements immediately (the lower loop) or by sending the prior slopes to a ANN predictor to extrapolate the current measurements (the upper loop). A reconstructor module (\texttt{Recon}) combines noisy slopes (either delayed, predictive, or delay-compensated) and control matrix generated during calibration to output DM commands, which are used by \texttt{DM} to generate the corrected phase. RMS error between the phase distortion and DM shape is then output as RMS WFE. 

Principal simulation parameters are listed in Table~\ref{tab:SimParas}. The configuration is adopted from CANARY low-order SCAO mode \citep{Tim2010}. We train the predictor under similar atmospheric and system conditions where it will be applied. The impact of the WFS SNR on ANN training and the predictor's robustness against changes in input statistics will be explored in Section~\ref{sec:results}. 

\subsection{ANN training}
\label{sec:training}
\subsubsection{Training data generation}
\label{subsec:trainingdata}
The wavefront sensing subsystem consisting of \texttt{Atmos} and \texttt{SH-WFS} modules is used to generate the first 100,000 training samples. Each sample is a time sequence of thirty 72-element vectors $(\textbf{s}_{1}, \textbf{s}_{2},...\ ,\textbf{s}_{30})$, with each vector, $\textbf{s}_{i}$, being the X and Y slope for each of the 36 subapertures. $(\textbf{s}_{1}, \textbf{s}_{2},...\ ,\textbf{s}_{29})$ will be ANN inputs sequentially during training, and $\textbf{s}_{30}$ will be the targeted output. Wind velocity corresponding to each sample is a random vector, with its magnitude uniformly sampled from the range 10 to 15 m/s and its direction uniformly sampled from the range 0 to \ang{360}. Wind velocity is constant within each sequence.

We then reverse each sequence to form the other half of the training set, with the last frame being the first and first being last. This corresponds to reversing the wind direction. We use this data augmentation approach to introduce variability in training data to improve model robustness. Resulting training input set and target set are tensors of shape ($2\times10^5$, 29, 72) and ($2\times10^5$, 72) respectively. The amount of training data is decided by trial and error to match both ANN architecture complexity and problem complexity to balance between training data fitting and model generalisation. No further training data pre-processing is implemented.

\begin{table}
\centering
 \caption{Principal parameters used with the Soapy SCAO simulation for ANN training and optimisation.}
 \label{tab:SimParas}
 \begin{tabular}{lll}
  \hline
  Module & Parameter & Value\\ [2pt]
  \hline
  System & Frequency & 150 Hz \\[2pt] 
  & Pure delay & 0.0067 s \\[2pt]
  & Throughput & 1 \\[2pt]
  & Gain & 1 \\[2pt]
  Atmosphere & \# of phase screens & 1\\[2pt]
   & Wind speed & 10-15 m/s\\[2pt]
   & Wind direction & 0-\ang{360}\\[2pt]
   & $r_0$ @ 500 nm & 0.16 m\\[2pt]
   & $L_0$ & 25 m\\[2pt]
  Telescope & Diameter & 4.2 m\\[2pt]
   & Central obscuration & 1.2 m\\[2pt]
  SH-WFS & GS magnitude & 10\\[2pt]
   & \# of subapertures & 7$\times$7 (36 active)\\[2pt]
   & Readout noise & 1 $e^-$ RMS\\[2pt]
   & Photon noise & True\\[2pt]
   & Wavelength & 600 nm \\[2pt]
   & Thresholding value & 0.1 \\[2pt]
  Piezo DM & \# of actuators & 8$\times$8\\[2pt]
  \hline
 \end{tabular}
\end{table}

\subsubsection{ANN training and optimisation}
\label{subsec:trainingandopt}
We use Keras \citep{Keras} library written in Python for ANN training. The ANN architecture consists of stacked LSTM cells and a final fully-connected (FC) output layer. The depth of neural networks is associated with the depth of representations that can be learnt \citep{Goodfellow2016}, thus the stacking of LSTM cells in our case.  

\begin{figure}
\centering
\includegraphics[width=\columnwidth]{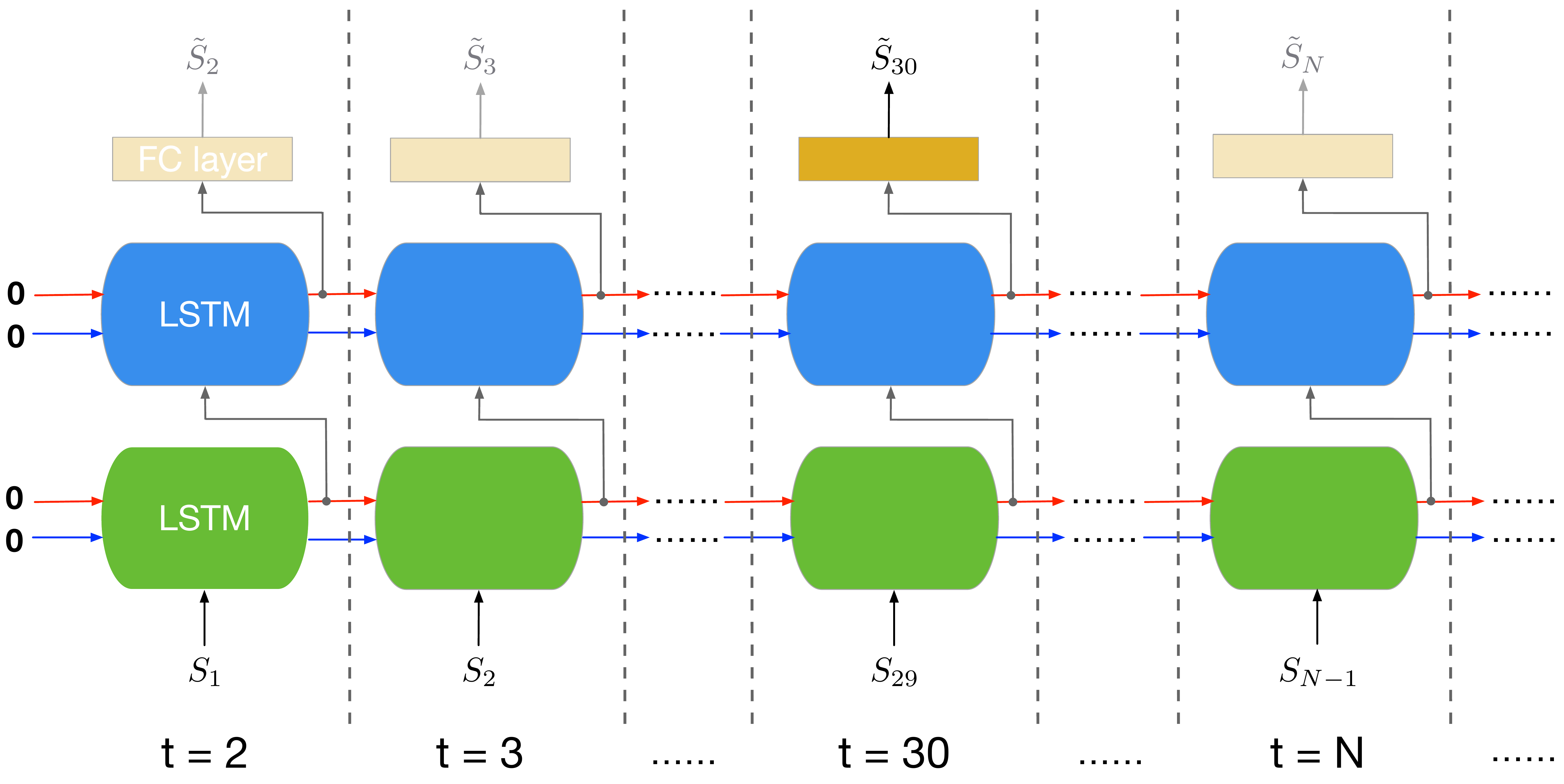}
\caption{The ANN predictor structure unrolled in time. The predictor can start predicting from the $2^{\textrm{nd}}$ time step, although initial predictions can be unstable and inaccurate due to limited temporal information. The two LSTM cells have the same inner structure, but different sets of parameters after training.} 
\label{fig:ANNfull}
\end{figure}

The ANN topology comprising two LSTM cells and a FC layer is shown in Fig.~\ref{fig:ANNfull}. The display is unrolled in time, which means all components in the same colour (or row) are duplicates in time and essentially identical to inputs at any time step. At each time step $t$ ($t\geq2$), the network can output a slope prediction ${\Tilde{\textbf{s}}}_{t}$ based on the current input $\textbf{s}_{t-1}$ and two state vectors, the cell state (also called the internal state) and the cell output (also called the hidden state). Both states are either initialised as all-zero vectors ($t=2$) or updated at each time step ($t>2$) using information in the input sequence so far. 

Parameters (also called trainable weights) of the network determine how inputs are processed mathematically layer by layer. ANN training is the process to optimise these parameters iteratively to minimise a training error. These parameters are initialised using a Gaussian distribution. 10\% dropout is deployed for each LSTM cell \citep{Gal2016}. 10\% training samples form a validation set. The remaining 90\% samples are randomly split into batches of size 128 before each epoch. The training error is mean squared error (MSE) between the targeted output ${\textbf{s}}_{30}$ and the actual output ${\Tilde{\textbf{s}}}_{30}$ evaluated and averaged on the current batch. The Adam optimisation algorithm is used to optimise the network parameters in a direction that minimises the training error \citep{Kingma2014}. It is a first-order gradient descent algorithm and features adaptive learning rate. During one epoch, every batch is evaluated once and the network parameters are updated accordingly multiple times. At the end of each epoch, the updated network is evaluated on the validation set. The initial learning rate is 1e-3. If MSE of the validation set shows no improvement for consecutive 10 epochs, the learning rate is reduced to its 1/5 unless reaching 1e-5. The reduced learning rate allows only small updates of the network parameters to prevent this optimisation process from early stagnation. Training is terminated after 40 epochs, at which point both training and validation errors have stagnated.

The ANN optimisation process, also called hyperparameter tuning, is coupled with ANN training. Hyperparameters determine either the structure of the network or the training process. These are fixed before training starts. We tune two hyperparameters that determine the physical capacity of the network: number of stacked LSTM cells (1 or 2) and length of output vectors of each LSTM cell (a random integer between 100 and 250, different for each cell). Every time a set of these two hyperparameters are chosen, the model is recompiled, re-initialised and re-trained as is described above. The model that achieves the lowest validation MSE at the end of the $40^{\textrm{th}}$ epoch is composed of two LSTM cells and a final FC layer (as is shown in Fig.~\ref{fig:ANNfull}). The output vector of the first LSTM cell has 247 elements and the second cell has 226 element. The resulting model has 761,000 trainable parameters in total. Breakdown of the number of floating point operations (FLOP) of the optimised ANN structure is shown in Table~\ref{tab:flops}. The resulting computational load is $2.3\times10^8$ FLOPS (FLOP per second) for the CANARY-scale $7\times7$ subaperture system operating at 150 Hz. 
\begin{table}
\centering
 \caption{Breakdown of computational load within the optimised ANN architecture.}
 \label{tab:flops}
 \begin{tabular}{llll}
  \hline
  Module & Input vector size & Output vector size & FLOPs\\
  \hline
  First LSTM & 72 & 247 & 630,344 \\ 
  Second LSTM & 247 & 226 & 855,184 \\
  FC & 226 & 72 & 32,544 \\
  \hline
  Total & & & 1,518,072 \\
  \hline
 \end{tabular}
\end{table}

\section{Results}
\label{sec:results}
After training, the optimised predictor is inserted between \texttt{SH-WFS} and \texttt{Recon} to form part of a predictive correction loop. From this stage, the parameters within the network are fixed and inputs are now processed in a deterministic way. We test the predictor's generalisation and extrapolation capabilities in five different scenarios:
\begin{enumerate}
    \item The predictor is tested on unseen data generated within the parameter boundaries used for the training regime.
    \item GS magnitude is increased from 10 (on which the predictor is trained) to 6, which increases the SNR of input slopes. In this scenario, we also investigate the SNR of training data on the predictor's performance. 
    \item A time-variant turbulence is considered by changing either the wind speed or the direction every 10 frames (15 Hz) after the predictor stabilises.
    \item A multi-layer turbulence is considered to test the predictor's ability to track multiple wind vectors.
    \item We extend our approach to account for a more realistic two-frame latency, where we trained a separate ANN predictor to predict two frames in advance directly, and compare that with applying the single-latency predictor twice.
\end{enumerate}
In most scenarios, statistics of the input slopes to the predictor are different to what was used during training. In each scenario, we use 1,000 test slope sequences each of 100 frames (0.67 s). We have found that our predictor will remain stable during a 2-minute period if both system and turbulence parameters remain unchanged, which is the case in most of the scenarios. Thus, in this paper we only include results obtained from 100-frame sequences to strengthen different aspects of the performance. The predictor's memory (both internal and hidden states) is zeroed before a new slope sequence. The predictor is expected to build up its memory and output stable predictions in 30 frames as the training is designed. Other simulation parameters are mostly the same as listed in Table~\ref{tab:SimParas}, unless stated otherwise.

\subsection{Performance within the ANN training regime}
\label{sec:scenOne}


\begin{figure}
    \centering
    \includegraphics[width=\columnwidth]{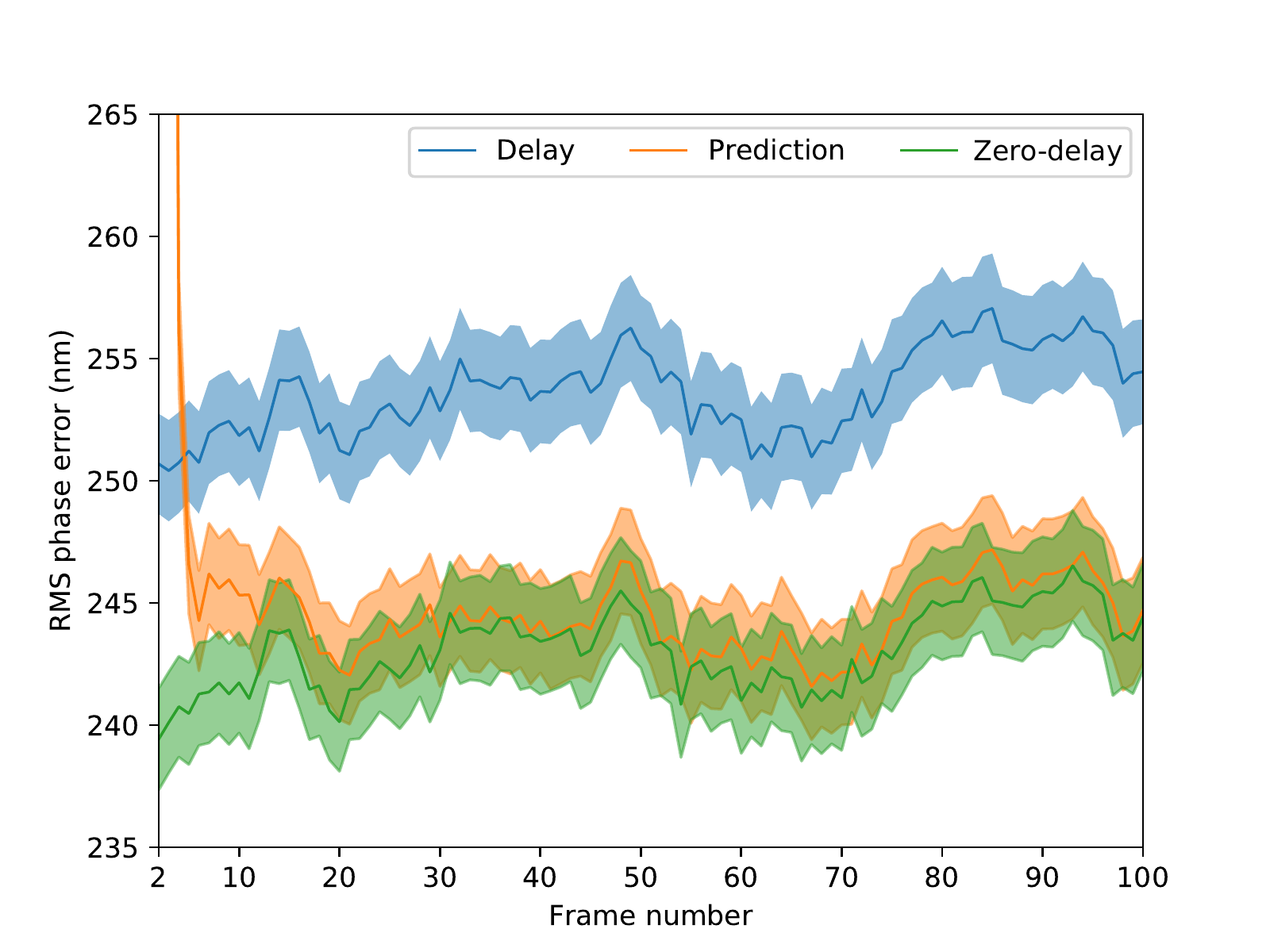}
    \caption{Mean RMS WFEs in an AO loop averaged across 1,000 test sequences. The GS used to generate test slopes has a magnitude of 10, which is the same as that for training. The predictor is tested within the training regime, though this test set had not been observed by the predictor before. Wind speed is 15 m/s in a single direction.}
    \label{fig:RMSE_10}
\end{figure}

\begin{figure}
    \centering
    \includegraphics[width=\columnwidth]{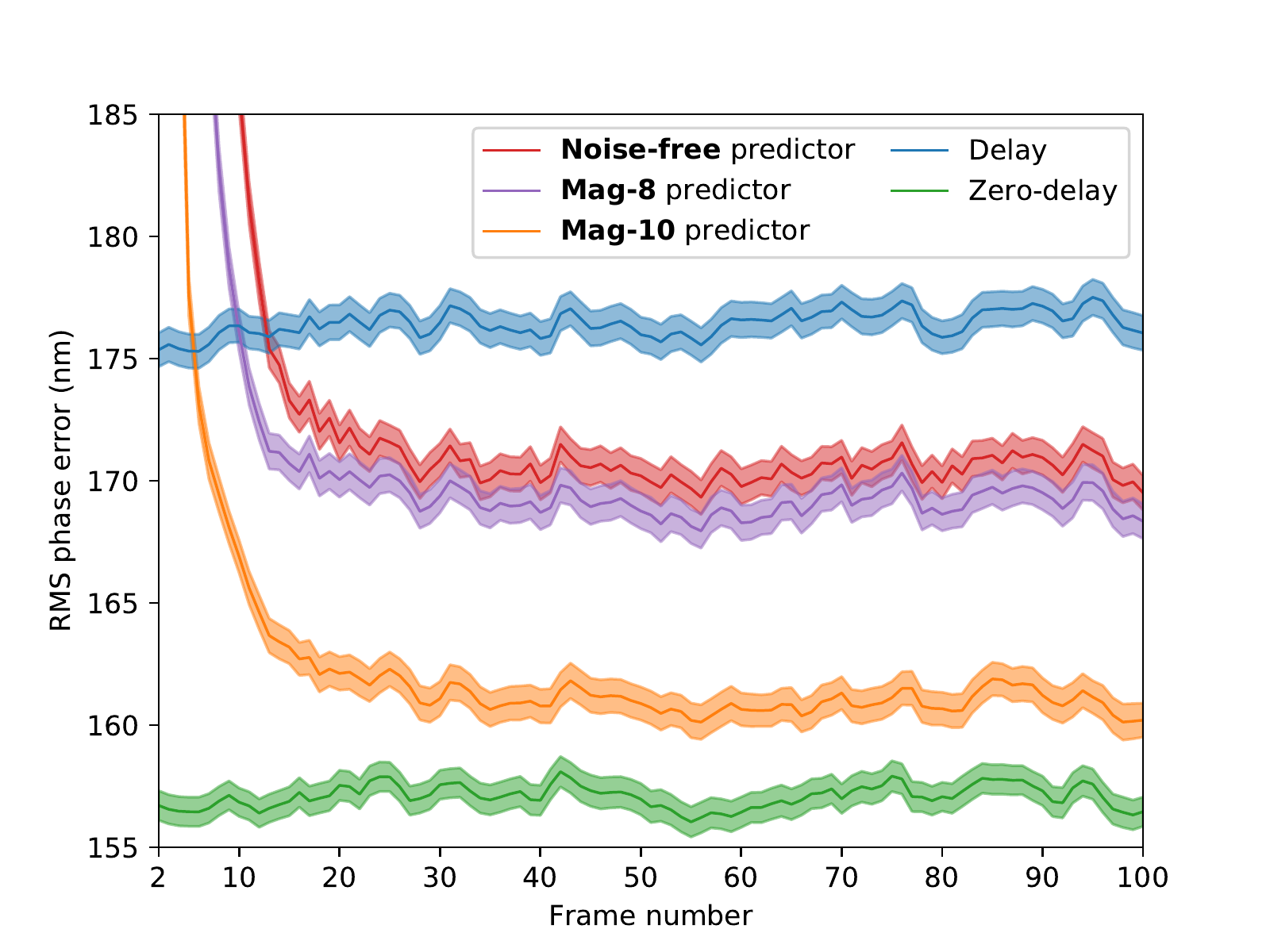}
    \caption{Mean RMS WFEs in an AO loop averaged across 1,000 test sequences. The GS used to generate test slopes has a magnitude of 6, which increases the SNR of inputs to the predictor that is trained with a GS of magnitude 10 (\textbf{Mag-10}) compared with during its training. Wind speed is 15 m/s in a single direction. We also compare \textbf{Mag-10} predictor using the same set of input slopes with another two predictors that are trained with GS magnitude 8 (\textbf{Mag-8}) and trained without WFS noise (\textbf{Noise-free})}.
    \label{fig:RMSE_6}
\end{figure}

Fig.~\ref{fig:RMSE_10} shows RMS WFEs averaged over 1,000 test atmospheric turbulence sequences. Shaded areas indicate the standard error of the mean RMS WFE \citep{Hughes2010}. The atmospheric statistics of this test set lie within the bounds of the training regime, though the test set did not form part of the training data set and had not been observed by the network before. Wind speed is 15 m/s in a single direction. All other simulation parameters are the same as during training, thus in this case the predictor is expected to reach its optimal performance. The predictor output stabilises after approximately 12 frames. The prediction is stable after this time span as the input statistics remain unchanged afterwards. This implies using shorter sequences for training and thus an alternative network that converges faster is possible. Mean RMS WFEs of the delayed, predictive and delay-compensated correction loop (averaged after the $12^{\textrm{th}}$ frame and across all sequences) are 253.9 nm, 244.3 nm and 243.4 nm respectively, showing an overall performance improvement brought by the predictor.

\subsection{Performance with varying WFS SNR}
\label{sec:scentwo}
In Fig.~\ref{fig:RMSE_6} we show the results from three ANNs when observing a bright guide star of magnitude 6. In addition to the ANN used in section \ref{sec:scenOne} that was trained on a guide star of magnitude 10, we include results from two networks have been trained at different signal to noise levels. These three predictors are denoted as \textbf{Mag-10}, \textbf{Mag-8} (trained with a GS of magnitude 8) and \textbf{Noise-free} (trained without WFS noise) respectively. The training procedure and other simulation parameters were the same as detailed in Section~\ref{sec:training}, except the thresholding value that was reduced to 0.02 for the \textbf{Mag-8} predictor, and 0 for the \textbf{Noise-free} predictor. The resulting ANN architectures and computational loads are listed in Table~\ref{tab:threemodels}. For each network, we see that prediction error decreases until the $20^{\textrm{th}}$ frame, after which the performance of each ANN stabilises. However, we note that the ANN trained with the lowest SNR performs far better than the ANNs trained in higher SNR regime and this behaviour was observed irrespective of guide star magnitude.

\begin{figure}
    \centering
    \includegraphics[width=\columnwidth]{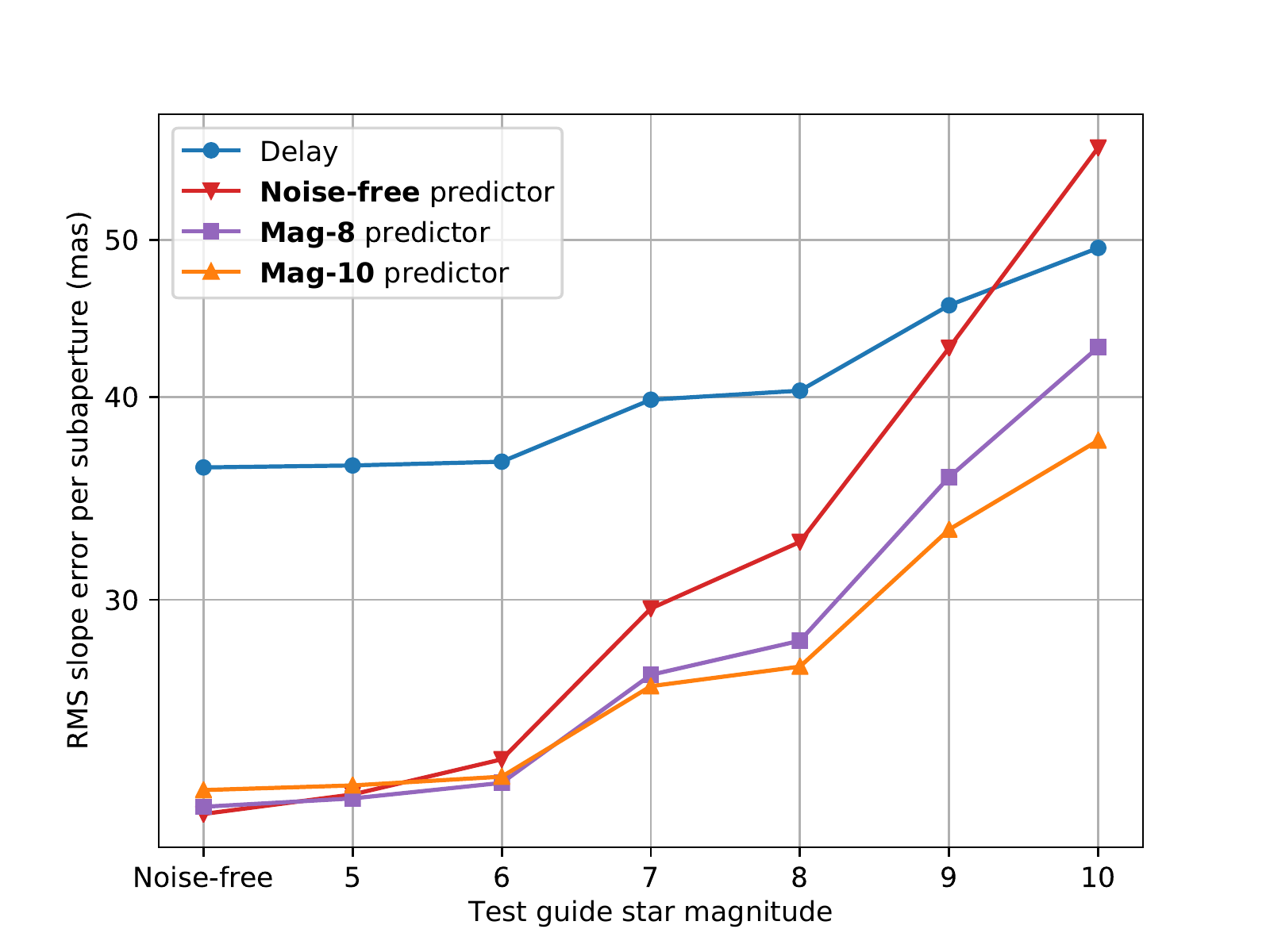}
    \caption{RMS slope error (mas) per subaperture compared with zero-delay measurements by \texttt{SH-WFS} as the WFS SNR varies. This quantity is the root of the ANN training metric. All predictors have lower errors around the corresponding training regimes.}
    \label{fig:slope_real}
\end{figure}

\begin{figure}
    \centering
    \includegraphics[width=\columnwidth]{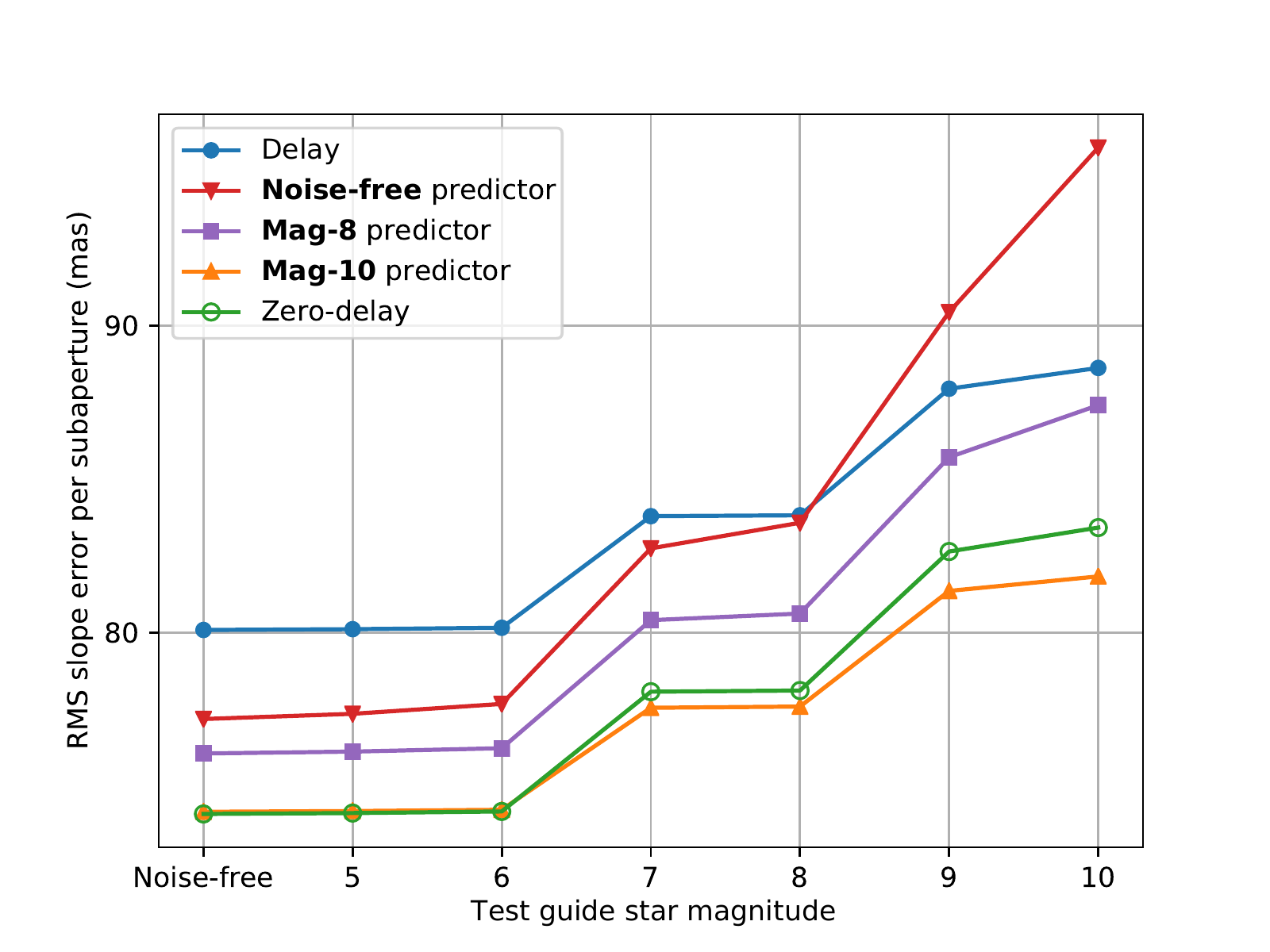}
    \caption{RMS slope error (mas) per subaperture with reference to measurements by the idealised WFS, which removes noise, aliasing and centroiding errors in the measurement of the first 36 Zernike orders compared with zero-delay measurements by \texttt{SH-WFS}. This along with Figs.~\ref{fig:RMSE_6} and \ref{fig:slope_real} demonstrates the filtering of aliasing and centroiding errors apart from the reduction in bandwidth error brought by \textbf{Mag-10} predictor.}
    \label{fig:slope_magic}
\end{figure}


\begin{table}
\centering
 \caption{Training conditions and structures of the three ANN predictors.}
 \label{tab:threemodels}
 \begin{tabular}{l|l|l|l}
  \hline
  ANN Predictor & \textbf{Mag-10} & \textbf{Mag-8} & \textbf{Noise-free}\\
  \hline
  GS magnitude during training & 10 & 8 & - \\ 
  WFS SNR during training & 17.6 & 52.5 & $\infty$ \\
  \hline
  \# of neurons of the first LSTM & 247 & 247 & 162 \\
  \# of neurons of the second LSTM & 226 & 203 & 114 \\
  \hline
  FLOPS @ 150Hz frame rate ($\times10^8$) & 2.3 & 2.1 & 0.9 \\
  \hline
 \end{tabular}
\end{table}


In Fig.~\ref{fig:slope_real} we show the RMS slope error (mas) per subaperture compared with zero-delay measurements as the WFS SNR changes. This quantity is the root of the ANN training metric. For each guide star magnitude, we generated 1,000 slope sequences each of 100 frames. Values shown in Fig.~\ref{fig:slope_real} give the mean slope error across all subapertures after the $30^{\textrm{th}}$ frame (by when all predictors have stabilised under all SNR conditions) in all sequences. All predictors have lower errors around the corresponding training regimes compared with slopes with one-frame delay, which shows the prediction power of ANN predictors of such type. In lower SNR regime (GS magnitude > 6), \textbf{Mag-10} predictor achieves lowest slope errors. However, at smaller GS magnitudes (GS magnitude $\leq$ 6) or in the noise-free condition, the performance of the \textbf{Mag-8} predictor is closer to that of the \textbf{Mag-10} predictor, rather than the \textbf{Noise-free} predictor. This is inconsistent with Fig.~\ref{fig:RMSE_6}. 

To assist understanding of this discrepancy in the brighter regime, we compare slope errors with reference to the slope measurements using an idealised WFS where centroiding and aliasing errors have been minimised. This WFS is defined as follows. The phase screen seen by the telescope aperture at each time step is firstly decomposed into its low- and high-order components,
\begin{equation}
    \boldsymbol{\phi} = \boldsymbol{\phi}_l + \boldsymbol{\phi}_h = \sum_{i=2}^{36}a_i\textbf{z}_i + \boldsymbol{\phi}_h,
\end{equation}
where $a_i=\textbf{z}_i^{\textrm{T}}\boldsymbol{\phi}$, $\textbf{z}_i$ is the $i^{\textrm{th}}$ Zernike term in Noll's notation \citep{Noll1976}. The number of Zernikes is chosen to match the structure of the subapertures. Measurements of $\boldsymbol{\phi}$ by the ideal WFS is computed as
\begin{equation}
    \textbf{s}^i \equiv \textbf{Da},
\end{equation}
where \textbf{D} is a perfectly calibrated interaction matrix and \textbf{a} is the Zernike vector ($a_2, a_3, ..., a_{36}$) given in Eq. 1. Within \textbf{D}, the slope measurements of each Zernike term are calculated directly from the corresponding high-resolution phase grid instead of from the WFS image plane. 

Fig.~\ref{fig:slope_magic} shows the RMS slope error with reference to the ideal WFS measurements in each correction loop. Errors in zero-delay measurements are non-zero due to aliasing, centroiding and noise errors compared to the ideal WFS. Among the three predictors, \textbf{Mag-10} predictor achieves significantly lower slope errors under all SNR conditions, which is now consistent with Fig.~\ref{fig:RMSE_6}. In addition we see that in low-SNR regimes, \textbf{Mag-10} predictor has even lower errors than zero-delay measurements. This implies that the reduction in WFE brought by \textbf{Mag-10} predictor also accounts for some aliasing and/or centroiding errors in addition to the reduction in bandwidth error. We think that this is due to being exposed to much lower SNR training data where the temporal correlations within the data are less obvious and noise terms must be learnt to be ignored. 

Prediction error $\sigma_{\textrm{pred}}$ is defined as the RMSE between WFEs in the predictive loop and in the zero-delay loop,
\begin{equation}
    \sigma_{\textrm{pred}} = \sqrt{{{\overline{WFE}}_{\textrm{pred}}}^2 - {{\overline{WFE}}_{\textrm{zero-delay}}}^2},
\label{eq:prederror}
\end{equation}
where ${\overline{WFE}}_{*}$ is the average after the $30^{\textrm{th}}$ frame and across all sequences. Bandwidth error is defined in a similar fashion,
\begin{equation}
    \sigma_{\textrm{BW}} = \sqrt{{{\overline{WFE}}_{\textrm{delay}}}^2 - {{\overline{WFE}}_{\textrm{zero-delay}}}^2}.
\label{eq:bwerror}
\end{equation}
$\sigma_{\textrm{pred}}$ of \textbf{Mag-10} predictor ranges from 40.0 nm to 19.9 nm, decreasing as the WFS SNR is lowered due to the increasing filtering of aliasing and/or centroiding errors. The mean value of bandwidth error across all SNR conditions is 78.3 nm, with a standard deviation of 4.4 nm. This quantity also decreases slightly as GS gets fainter, due to the increasing correlation between bandwidth error and noise error. 

In the following three scenarios, we show the results obtained with our optimal \textbf{Mag-10} predictor only. We also use a brighter guide star of magnitude 6 to reduce the performance variations brought by wavefront sensor noise.

\subsection{Performance with time-variant wind velocity}
\label{sec:scenthree}
In the above scenarios, we have assumed stationary turbulence. In this section, we demonstrate the agility and robustness of our ANN predictor against fluctuations in wind velocity. 

\begin{figure*}
\begin{minipage}{0.48\textwidth}
\centering
\includegraphics[width=\columnwidth]{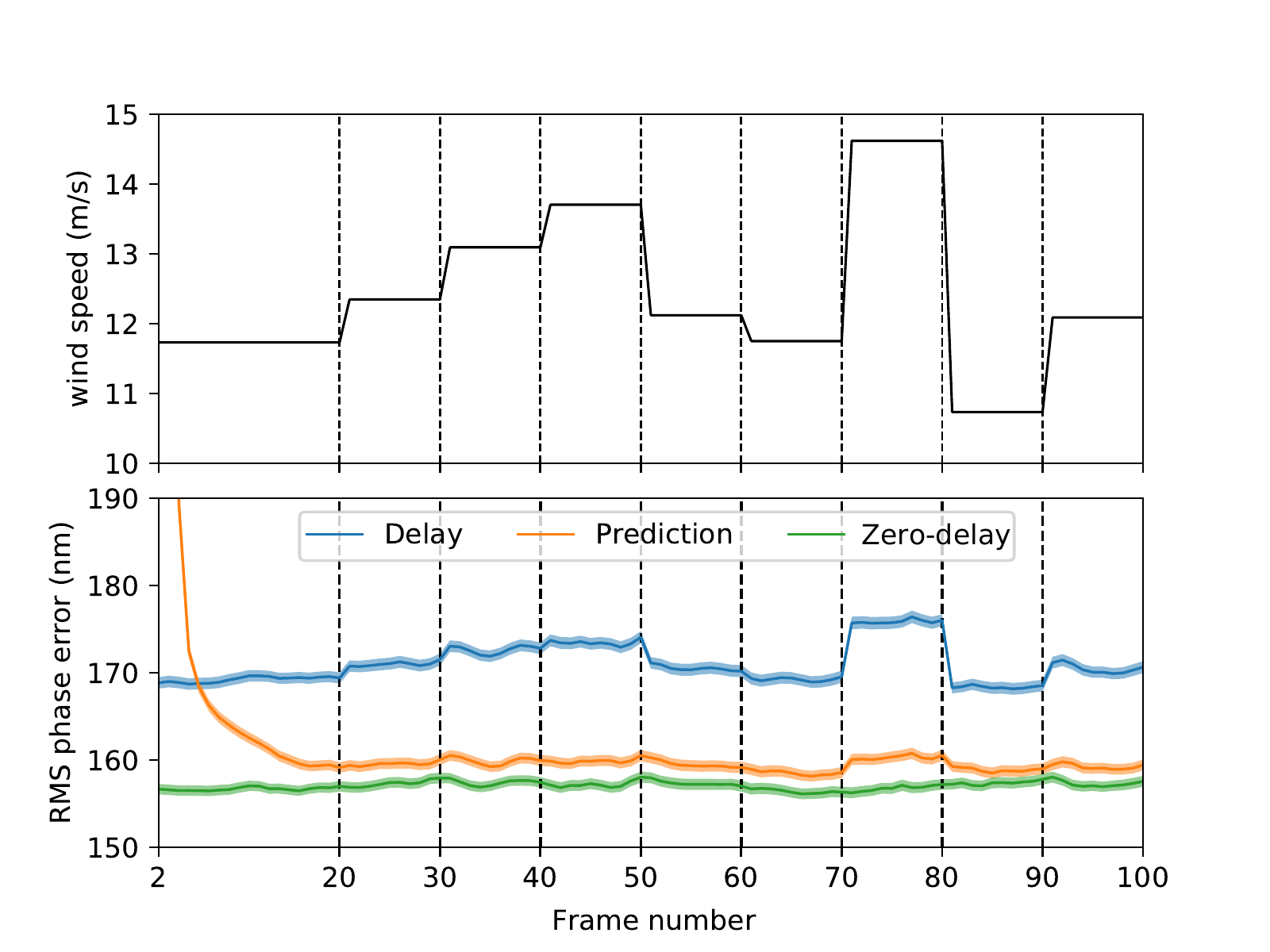}
\caption{Robustness of the predictor against wind speed fluctuations between 10 and 15 m/s every 10 frames. Wind direction is 0 degree. Guide star magnitude is 6.}
\label{fig:vchange_small}
\end{minipage}\hfill
\begin{minipage}{0.48\textwidth}
\centering
\includegraphics[width=\columnwidth]{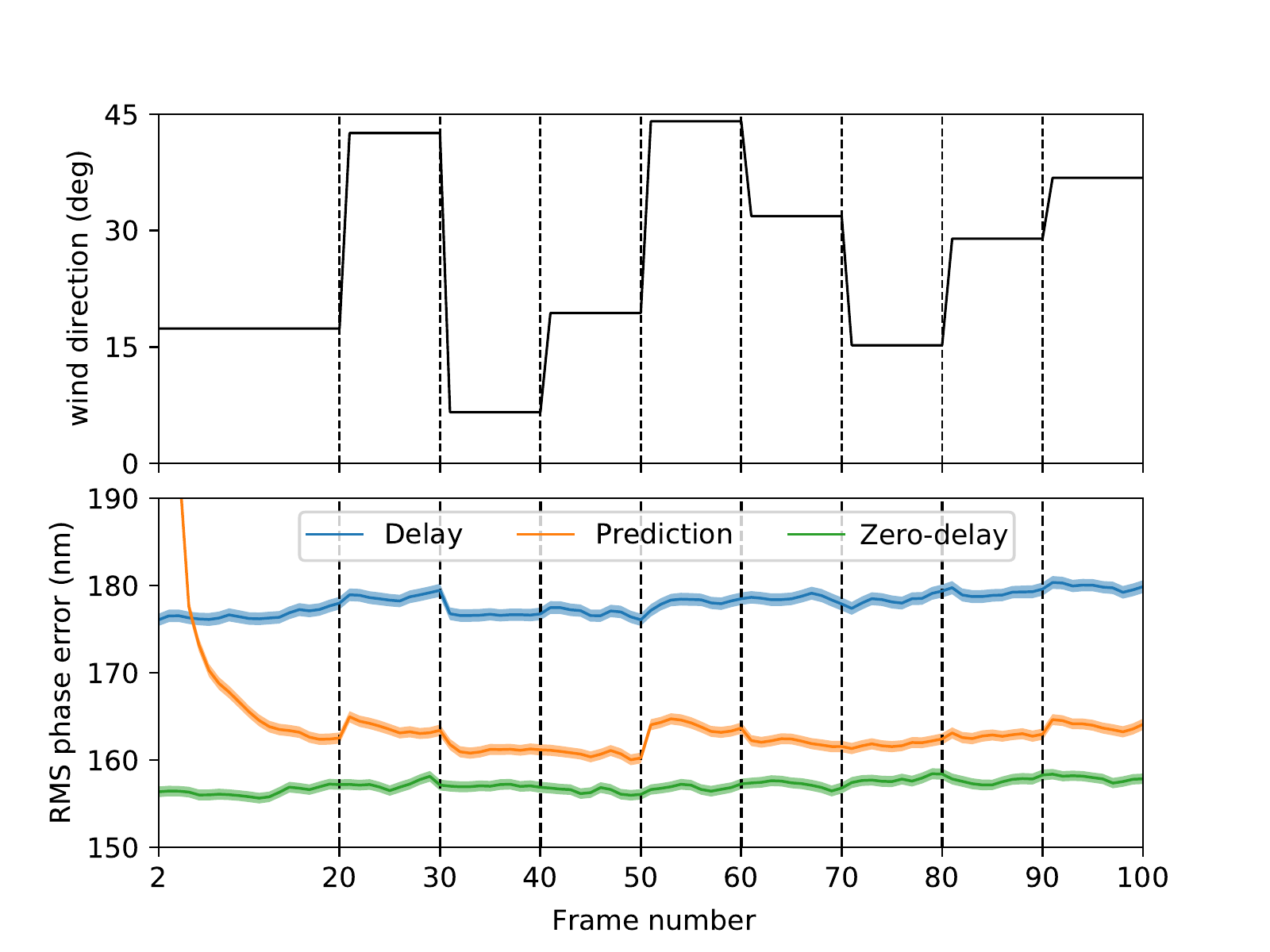}
\caption{Robustness of the predictor against wind direction fluctuations between 0 and 45 degrees every 10 frames. Wind speed is 15 m/s. Guide star magnitude is 6.} 
\label{fig:vchange_dir}
\end{minipage}
\end{figure*}

Here we use a synthetic wind speed sequence (upper panel in Fig.~\ref{fig:vchange_small}) in a relatively short time scale of 100 consecutive WFS frames (0.67 s). Wind speed changes every 10 frames (15 Hz) within 10 and 15 m/s after the first 20 frames during which time the predictor stabilises. This fluctuation is reflected in the dynamics of the delayed correction, as a faster translation of the phase screen induces increased phase variations between adjacent frames under frozen flow. 

Fig.~\ref{fig:vchange_dir} demonstrates robustness of the predictor against wind direction fluctuations between 0 and 45 degrees every 10 frames (upper panel). This corresponds to a maximum instantaneous change of 8.4 m/s in wind speed along a single direction. 

Recently \citet{Kooten19} have used typical wind profiles from the Thirty Metre Telescope (TMT) site to demonstrate effects of wind velocity variations in a data-driven linear minimum mean square error (LMMSE) predictor over a period of 5 seconds in numerical simulations. Wind data are linearly interpolated to system frequency to allow for per-frame fluctuation. Two adaptive variations, resetting-batch LMMSE and forgetting LMMSE, along with LMMSE were tested. Compared with these linear predictors, variances in WFEs of the ANN predictive loop before and after the disturbance are on the same order as that of the delay-compensated loop. This robustness can be explained as the ANN predictor is allowed to use more spatial and temporal information when making inferences. Furthermore, the updating and forgetting mechanisms of our predictor are not fixed, but can constantly self-adjust according to the inputs, which by design allows for more flexible control on data flow. 


\subsection{Performance with multi-layer turbulence}
\label{sec:scenfour}
Though we train the predictor with a single turbulence layer, there usually exists several layers at high altitudes in addition to a strong ground layer \citep{Ollie18,Ollie19}. It is thus meaningful to test the predictor's sensitivity to multiple layers moving with different velocities. 

Here we show the results obtained 
with ESO (European Southern Observatory) median 35-layer profile. $r_0$ is 0.157 m, slightly worse than during ANN training. We generated 1,000 slope sequences each of 100 frames with this profile. For comparison, we also generated the same amount of test data of a single ground layer and of a four-layer profile (detailed in Table~\ref{tab:multilayer}), both moving at 9.21 m/s (slightly slower than the training range), which is equivalent to the dynamics of the 35-layer profile.

\begin{figure*}
\begin{minipage}{0.48\textwidth}
\centering
\includegraphics[width=\columnwidth]{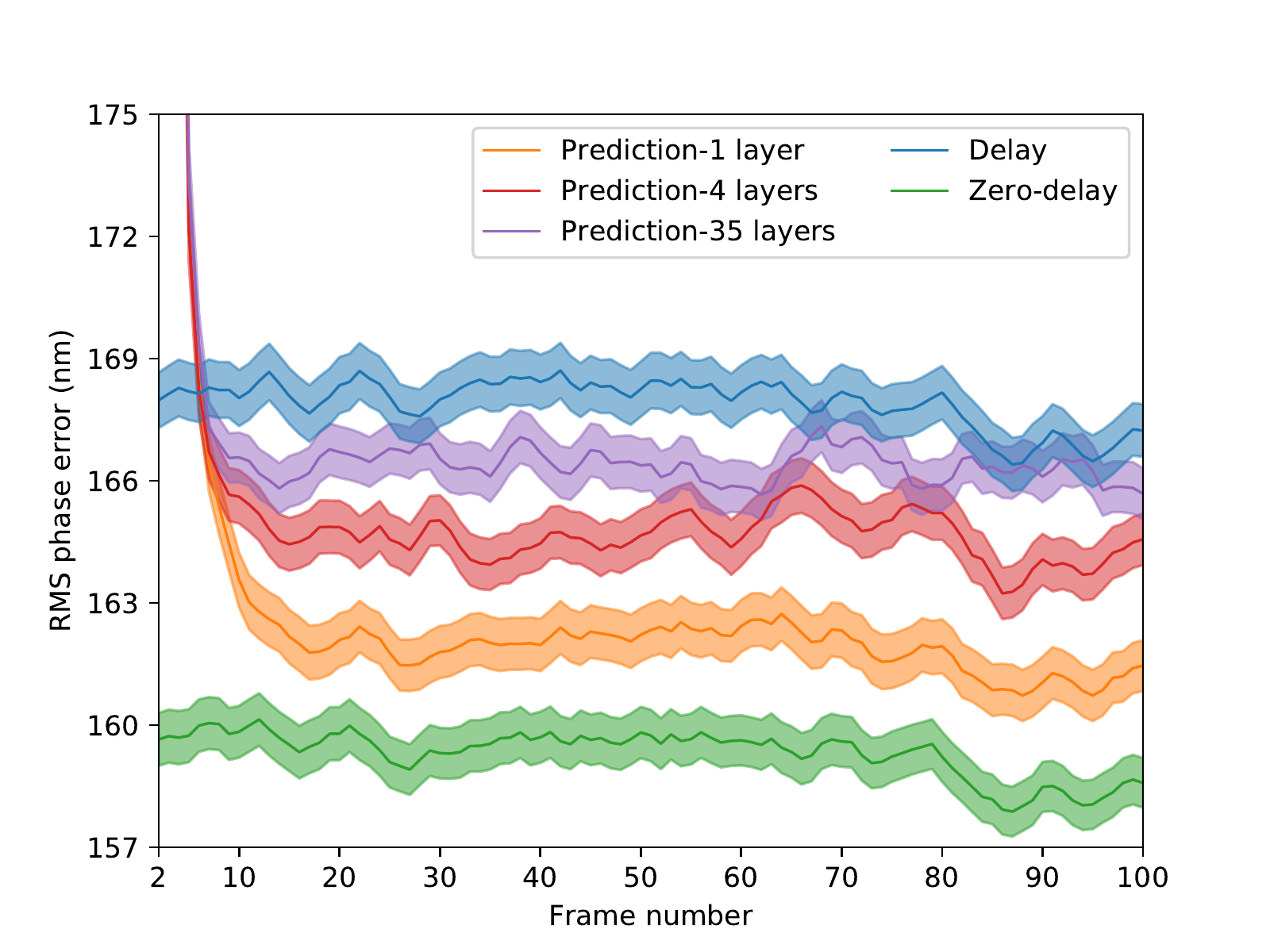}
\caption{ANN performance with multiple turbulence layers moving along different directions. Wind speeds of either the 1- or 4-layer profile are scaled to maintain the same dynamics as that of the 35-layer profile. $r_0$ is 0.157 m.}
\label{fig:multilayer_alldir}
\end{minipage}\hfill
\begin{minipage}{0.48\textwidth}
\centering
\includegraphics[width=\columnwidth]{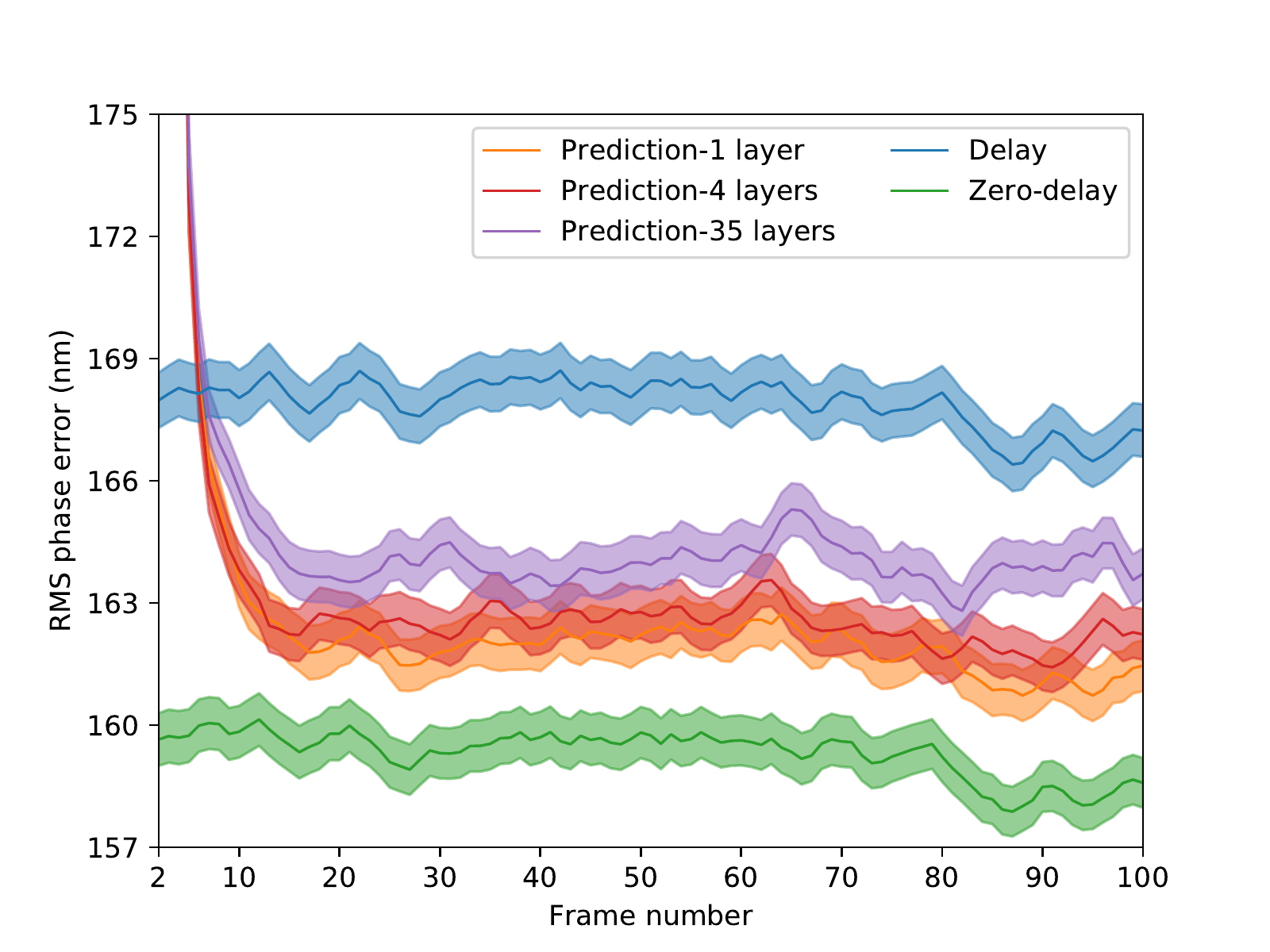}
\caption{ANN performance with multiple turbulence layers moving along the same direction. Compared with Fig.~\ref{fig:multilayer_alldir}, the ANN performance suffers from the increased number of wind vectors, but mainly from the variety among those vectors.}
\label{fig:multilayer_zerodir}
\end{minipage}
\end{figure*}

Fig.~\ref{fig:multilayer_alldir} shows residual WFEs when wind vectors of multi-layer profiles (either the 4-layer or the 35-layer) move in different directions. For the 35-layer profile, the moving direction of each layer is a random integer between 0 and 360 degrees. For the 4-layer profile, wind directions are listed in Table~\ref{tab:multilayer}. The delayed and the delay-compensated correction loops behave similarly regardless of the number of layers, thus only values obtained from the single-layer profile are shown here. Mean RMS WFEs of the delayed, 35-layer predictive, 4-layer predictive, 1-layer predictive and delay-compensated correction loop after the $20^{\textrm{th}}$ frame are 167.9 nm, 166.4 nm, 164.6 nm, 161.9 nm and 159.2 nm respectively. 

Fig.~\ref{fig:multilayer_zerodir} shows improved ANN performance when all layers in either multi-layer profile move in the same direction (wind speeds are the same as used in Fig.~\ref{fig:multilayer_alldir}). Mean RMS WFE of the 35-layer predictive loop decreases to 164.0 nm, slightly better than the 4-layer predictive loop when wind vectors are largely distinct from each other. Mean RMS WFE of the 4-layer predictive loop decreases to 162.4 nm, approaching that of the 1-layer predictive loop.

We think that the wind directions adopted represent two extreme conditions, and that performance with real turbulence profiles would fall within these two cases. These results show that the predictor trained on a single layer frozen-flow conditions is capable of providing performance improvement even when complex profiles with random wind directions are encountered.

\begin{table}
    \label{tab:multilayer}
    \centering
    \caption{Four-layer turbulence profile used within test dataset. $r_0$ is 0.157 m. $L_0$ is 25 m. Two sets of wind directions corresponding to Figs.~\ref{fig:multilayer_alldir} and \ref{fig:multilayer_zerodir} respectively are examined.}
    \begin{tabular}{l|l|l|l|l}
    \hline
        & Layer 1 & Layer 2 & Layer 3 & Layer 4 \\
    \hline
        Height (m) & 0 & 4000 & 10000 & 15500 \\
    \hline
        Relative strength & 0.65 & 0.15 & 0.10 & 0.10 \\
    \hline
        Wind speed (m/s) & 7.6 & 9.5 & 11.4 & 15.2 \\
    \hline
        \multirow{2}{*}{Wind direction (degrees)} & 0 & 330 & 135 & 240 \\ \cline{2-5} & 0 & 0 & 0 & 0 \\
    \hline
    \end{tabular}
    
\end{table}

\subsection{Performance with two-frame latency}
So far we have considered only single-frame delay in an AO loop, where we have accounted for WFS integration time only but ignored the time taken for real-time processing and the update of the surface shape of the DM.

\begin{figure}
    \centering
    \includegraphics[width=\columnwidth]{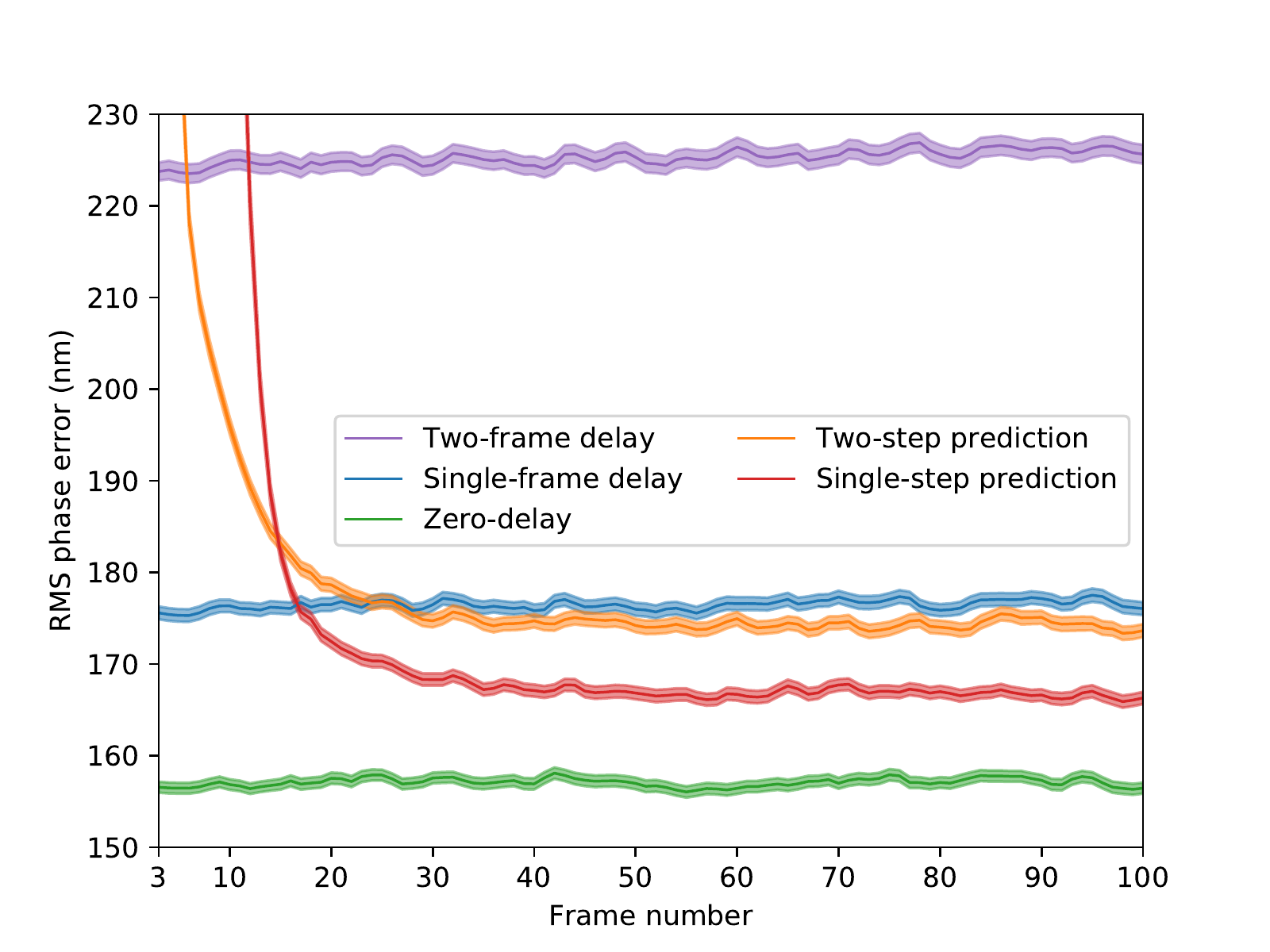}
    \caption{In a simulated system with a two-frame latency, the methodology adopted for the single-latency prediction is extended to training a separate ANN predictor (single-step prediction). In this case, the single-latency predictor can also be used twice (two-step prediction), albeit with worse performance. Both predictors improve the system performance significantly. The blue line representing the single-frame delay performance is the same as that shown in Fig.~\ref{fig:RMSE_6}, and is depicted here for comparison with the two-frame delay performance. Guide star magnitude for test is 6. Wind speed is 15 m/s along a single direction.}
    \label{fig:twofrms}
\end{figure}

In Fig.~\ref{fig:twofrms} we show the ANN performance when a more realistic loop delay of two frames is considered. We trained a separate ANN that was designed to predict two frames in advance in a single step. The training dataset described in Section~\ref{subsec:trainingdata} was re-utilised in the way that $(\textbf{s}_{1}, \textbf{s}_{2},...\ ,\textbf{s}_{28})$ in each sequence is the ANN input and $\textbf{s}_{30}$ is the training target. The training and hyperparameter searching setup follows that described in Section~\ref{subsec:trainingandopt}. The resulting network comprises two stacked LSTM cells and a final FC layer. The output vector sizes of the two LSTMs are 122 and 171 respectively, with a computational load of $0.9\times10^8$ FLOPS. The resulting mean RMS WFE of this single-step predictive loop after the $30{^\textrm{th}}$ frame is significantly reduced to 166.9 nm, compared with 225.6 and 157.1 nm of the two-frame delayed and zero-delayed loop respectively.

As a comparison, the single-frame predictor was also used twice to provide a two-frame prediction: first, the measured $(\textbf{s}_{1}, \textbf{s}_{2},...\ ,\textbf{s}_{t})$ ($t\geq2$) is fed into the predictor to generate the predicted $\Tilde{\textbf{s}}_{t+1}$ as it was designed; second, $\Tilde{\textbf{s}}_{t+1}$ is treated as its truth value $\textbf{s}_{t+1}$ and forms part of the ANN input vector $(\textbf{s}_{1}, \textbf{s}_{2},...\ ,\textbf{s}_{t}, \Tilde{\textbf{s}}_{t+1})$, which is then used to generate $\Tilde{\textbf{s}}_{t+2}$. This resulted in a WFE of 174.4 nm, worse than the two frame prediction, however still significantly better than the two-frame delay.


\section{On-sky implementation}
The results presented within this paper are based on simulations, which do not consider many of the practical issues relating to the implementation of the LSTM architecture within a real AO system. In this section we discuss issues relating to calibration and control within a real system.

The training method presented here uses simulated Shack-Hartmann WFS slope data but could be applied to real data from any WFS data. However, the sensitivity of the ANN performance to the training regime and the requirement for a large number of wind velocities required for a robust ANN training means that the collection of a real WFS datasets may take a significant amount of time. It may therefore be best to initially train in simulation and convert real WFS slopes to ensure that the subaperture geometry and pixel scale matches that encoded within the ANN. Unlike other ANN approaches proposed for multiple guide star AO \citep{Osborn2012}, the single WFS LSTM ANN does not require retraining for different targets, greatly reducing the calibration overhead of implementation within a real system.

The ANN predictor proposed here may not be applicable to all closed-loop AO systems where imperfect POLC (pseudo open loop control) can introduce additional noise terms within the system, affecting performance and loop stability \citep{Gilles05}. To adapt the training regime here to closed-loop operation the training dataset would have to be expanded to include the range of potential closed-loop gain values. This will increase both training time and the size of the resulting ANN, with no guarantee that the resulting ANN would be more resistant to the errors that can affect POLC stability such as misalignments and open-loop DM errors. 

An on-sky implementation of the ANN presented here requires an additional processing step for each WFS before reconstruction within the system. The system simulated here uses a $7\times7$ subaperture Shack-Hartmann system that we selected such that can be rapidly trained and tested. Furthermore, wind profiles can be recovered from recorded off-axis WFS data of the CANARY demonstrator \citep{Dougie19}. By matching the configuration of CANARY, future comparison of multi-layer predictions using real data and in simulation is feasible. Extending beyond this low-order system is possible but implies additional training time and an increase in real-time computational load. Due to the hyperparameter tuning approach adopted here, the precise computational load of a higher-order system cannot be easily predicted, but implementation of this approach within any existing astronomical non-XAO system is feasible using existing hardware. There do however exist possibilities to reduce the computational load, including operating in actuator space where computational load is lower \citep{Basden2019}, or taking advantage of the sparsity of the ANN. The technique proposed here inherently scales to multiple guide star systems through parallelism.

\section{Conclusions}
We have shown in extensive numerical simulations the potential of artificial neural networks as a nonlinear framework for wavefront prediction. The memory elements within the LSTM network give it the ability to learn information such as wind velocity vectors from the data and to use that information in its prediction. The fluid nature of the memory allows the network to adapt to changes in such information without user tuning. 

The residual wavefront error of the simulated $7\times7$ subaperture SCAO system with one frame delay improves significantly after the predictor is incorporated irrespective of guide star magnitude and wind velocity. In addition to accurately predicting the wavefront we have also provided evidence that the ANN predictor also compensates for some centroiding and/or aliasing errors that can be temporally filtered from the wavefront. This behaviour however is dependent on the ANN training regime and was only observed when the system was trained on a low SNR $10{^\textrm{th}}$ magnitude guide star. The selection of the training regime has the greatest impact on the performance of the ANN prediction. 

We have shown that the ANN predictor is robust to changes in wind velocity on sub-second timescales, and that the ANN approach taken within this paper is transferable to systems with a two-frame delay. The ANN predictor trained on a single atmospheric turbulence layer is also capable of predicting under more complex conditions with multiple layers with independent wind vectors, albeit with reduced performance. Whilst we believe it is likely that a more realistic multi-layer training environment and/or the use of multiple wavefront sensors to allow identification of layer altitudes will improve ANN performance on multi-layer turbulence, this is subject to further study. Our next steps will be to investigate ANN performance on recorded CANARY data to investigate ANN stability and training in a real-world system.


\section*{Acknowledgements}
The authors gratefully acknowledge James Osborn from CfAI, Durham University for his valuable suggestions regarding the manuscript. We thank sincerely Ollie Farley from CfAI for the thought-provoking discussions with him. We sincerely appreciate the comments from the reviewer, which helped us improve the paper. X. Liu is funded by Durham University through their doctoral scholarship program and China Scholarship Council. CGG and JDCJ acknowledge financial support from the I+D 2017 project AYA2017-89121-P and JDCJ acknowledges support from the European Union’s Horizon 2020 research and innovation programme under the H2020-INFRAIA-2018-2020 grant agreement No 210489629. 







\appendix



\bsp	
\label{lastpage}
\end{document}